\documentclass[10pt]{iopart} 

\usepackage{epsfig,graphicx}
\usepackage{subfigure,xcolor}

\def\msun{M_\odot}
\def\rsun{R_\odot}
\def\mbh{M_\bullet}
\def\kms{{\rm km~s^{-1}}}
\def\Sgr{{Sgr A$^*$ }}

\def\mnras{{\it Mon. Not. Roy. Astron. Soc. }}

\def\apj{{\it Astrophys. J.}}
\def\apjs{{\it Astrophys. J. Sup. Ser.}}
\def\apjl{{\it Astrophys. J. Lett.}}
\def\nat{{\it Nature}}
\def\aap{{\it Astron. \& Astrophys.}}

\def\aj{{\it Astron. J.}}
\def\araa{{\it Annu. Rev. Astron. Astrophys.}}
\def\na{{\it New Astron.}}

\def\keywords#1{\vspace{10pt}
     \begin{indented}
     \item[]\rm Keywords: #1\par
     \end{indented}}

\newcommand\arcsec{\mbox{$^{\prime\prime}$}}%

\begin{document}

\title{Sculpting the Stellar Cusp in the Galactic Center}

\author{Xian Chen$^1$ \& Pau Amaro-Seoane$^1$}

\address{$^1$\ Max Planck Institut f\"ur Gravitationsphysik
(Albert-Einstein-Institut), Am M\"{u}hlenberg 1, D-14476 Potsdam, Germany.}

\begin{abstract}

Observations of the innermost parsec surrounding Sgr A* ---the supermassive
black hole in the center of our Galaxy--- have revealed a diversity of
structures whose existence and characteristics apparently defy the fundamental
principles of dynamics.  In this article, we review the challenges to the
dynamics theories that have been brought forth in the past two decades by the
observations of the Galactic center (GC). We outline the theoretical framework
that has been developed to reconcile the discrepancies between the theoretical
predictions and the observational results. In particular, we highlight the role
of the recently discovered sub-parsec stellar disk in determining the dynamics
and resolving the inconsistencies. We also discuss the implications for the
recent activity of Sgr A*. 

\end{abstract}




\section{Introduction}\label{sec:intro}

Since the late 1970s, there was already circumstantial evidence pointing to
the existence of a supermassive black hole (SMBH) in the dynamical center of
our Galaxy \cite{wollman77,lacy80}. Later observations pinpointed the location
of the SMBH at \Sgr, the compact radio object now defining the Galactic center
(GC) \cite{lo87}. The radiative characteristics of \Sgr suggest that it is
fueled by plasma at an extremely low rate \cite{narayan95,narayan98,yuan02}.
These early discoveries are consonant with the hypothesis that SMBHs were born
and growing in the early universe, manifesting themselves as active galactic
nuclei (AGN) \cite{schmidt63,LyndenBell69}, but today they are mostly ``dead'',
lurking in the nuclei of quiescent galaxies \cite{soltan82,yu02qso}.

More solid evidence of the SMBH in the GC comes from stellar dynamics.  In the
past two decades, the advent of the adaptive optics and integral field
spectroscopy allows us to detect stars in the close vicinity of \Sgr and
monitor their orbits with high precision
\cite{ghez00,eckart02,eis05,ghez05,gillessen09}).  So far, dozens of B-type
stars (so called ``S-star cluster'', or ``S-cluster'') have been discovered in
a region as close as $<1\arcsec$ ($1\arcsec\simeq0.04~{\rm pc}\simeq0.13~{\rm
light~year}$ at the GC) from \Sgr \cite{schodel03,eis05,ghez05,gillessen09}.
Their near-Keplarian orbits have provided the hitherto most stringent
constraint on the mass of a SMBH, which is $\mbh\simeq4.3\times10^6~\msun$
(e.g. \cite{gillessen09}). Such a tight constraint firmly anchors our Milky Way
(MW) toward the lower end of the correlation, shared among the galaxies in the
local universe, between the masses of their central SMBHs and the properties
(such as mass and stellar velocity dispersion) of their stellar components
\cite{kormendy13}.

Meanwhile, photometric and kinematic observations of the stars in a larger area
surrounding \Sgr have revealed a rich family of structures (see \cite{genzel10}
for a review). Now it is clear that the S-cluster is the innermost part of a
young star cluster, which is composed of mainly B-stars (each has a mass of
$7-20~\msun$ and a typical age of $\sim10-10^2$ Myrs) and extending
continuously out to a distance of several pc from \Sgr
\cite{schodel07,buchholz09,do09,do13,madigan14}. This cluster spatially
coincides with an old (several Gyr) population of red giants (RGs), whose
surface density exhibits a characteristic ``core-like'' profile,  such that at
the innermost $0.5$ pc of the cluster, the surface density of the RGs becomes
constant or possibly decreasing toward the central SMBH
\cite{buchholz09,do09,merritt10,do13}. The young B-stars and the old RGs in the
GC constitute the nuclear star cluster (NSC) of the MW. There is another
dynamically distinct structure centered on \Sgr and spanning a radial range of
$(0.04-0.5)$ pc, known as the ``mini disk'', which consists of $O(10^2)$ young
($6\pm2$ Myr) and massive ($>20~\msun$) Wolf-Rayet (WR) and O-type
stars, in a configuration of a (possibly two) mildly thick disk(s).  Apart from
these discoveries, it is likely that so far we have spotted only the tip of the
iceberg -- the vast majority of the fainter, lower-mass stars may be ``hiding''
behind the bright glow in this very crowded field.

The rich phenomena unveiled in the GC offer a starting point for us to test our
various hypotheses of the stellar and gaseous processes that may be responsible
for the formation and evolution of SMBHs in general \cite{melia01,alexander05}.
However, when contrasting the observations with the theories, many
inconsistencies occur \cite{alexander11}.  For example, the core-like
distribution of RGs around \Sgr contradicts the longstanding hypothesis that
stars in the potential well of a SMBH will relax due to two-body interactions
to form a cusp whose stellar density should rise steeply toward the central
SMBH (see \cite{peebles72,BahcallWolf76,shapiro76,lightman77} for the original
idea and \cite{pau12} for a recent review). Furthermore, the existence of the
S-cluster so close to \Sgr imposes a paradox \cite{morris93,ghez03}: the stars
there cannot form in situ due to the violent environment, therefore they must
be old so they have enough time to migrate in from outside due to dynamical
friction, but S-stars are young. These discrepancies indicate that our
comprehension of the stellar and gaseous processes close to a SMBH is far from
complete.

During the past two decades, in the course of reconciling our dynamical model
of the GC with the observations, new physical processes have been discovered
and new theories have been formulated. These progresses have greatly enriched
our understanding of the relationship between SMBHs and their close
environments.  In this contribution to the special issue, we will review the
challenges we have been facing, as well as the potential solutions. In
particular, we will highlight the recent discovery of the impact of the mini
disk on the dynamics in the GC and its role in reconciling theories with
observations.

The article is organized as follows.  In section~\ref{sec:RGs} we describe the
``missing RG problem'' and review the conventional solutions as well as their
limits.  We then introduce the most recent idea that it is caused by the
repeated collisions of the RGs with the gaseous clumps in the fragmenting past
of the mini disk.  In Section~\ref{sec:Sstars}, we outline the solutions to the
``paradox of youth'' imposed by the S-cluster.  We also highlight the  impact
of the mini disk on the dynamical evolutions of the S-stars.  The same
dynamical effect of the disk offers a potential solution to the conundrum of
``inversed mass segregation'', that is manifested in the spatial distribution
of young stars.  This idea is elaborated in Section~\ref{sec:WROstars}.
Finally, in Section~\ref{sec:outlook}, we summarize, and generalize the
conclusions to other quiescent galaxies.

\section{Culprit for the Missing Red Giants}\label{sec:RGs}

\subsection{Is there a Bahcall-Wolf Cusp in the GC?}\label{sec:BWcusp}

The existence of a SMBH differentiates a region around it where the
gravitational potential is dominated by that of the black hole (BH). In the GC,
this region has a characteristic size of

\begin{equation}
r_h\equiv\frac{G\mbh}{\sigma^2}\simeq1.7~{\rm pc}
\left(\frac{\mbh}{4\times10^6~\msun}\right)\left(\frac{\sigma}{100~{\rm km~s^{-1}}}\right)^{-2},
\end{equation}

\noindent
where $\sigma$ characterizes the velocity dispersion of the surrounding stars.

Inside the sphere of this ``influence radius'' ($r_h$), a star revolves around
the central SMBH on a near-Keplerian orbit, whose pericenter precesses in the
orbital plane, at a rate determined by the gravitational potential of the NSC
(Newtonian precession) as well as by the general-relativistic effect of the
SMBH (GR precession). The star is unperturbed most of the time. But
occasionally, it will encounter another star, exchanging energy and angular
momentum with the interloper, which alters the orbits of both stars. Such
``two-body interactions'' redistribute energy and angular momentum between
stars, a dynamical process called ``relaxation''. The equilibrium outcome, as
has been predicted four decades ago, is a stellar cusp characterized by a
steep-rising density profile toward the central SMBH, $\rho\propto r^{-7/4}$,
also known as the ``Bahcall-Wolf profile''
\cite{peebles72,BahcallWolf76,shapiro76,lightman77}.

A number of studies more specific to the conditions in our Galaxy argued that a
Bahcall-Wolf cusp is a robust outcome
(\cite{freitag06,hopman06ms,preto10,pau11,gualandris12}, and see
\cite{antonini12df} for alternative).  However, this prediction apparently
contradicts the observational fact that the RGs in the GC, whose ages are
several Gyr, do not show a cuspy distribution, but instead they exhibit a flat,
core-like profile at $r<0.5$ pc
\cite{genzel96,buchholz09,do09,merritt10,yusef11}. This discrepancy is called
``the problem of missing RGs''. A number of solutions have been proposed.  Most
of them spring from two opposite standpoints: whether or not the RGs are
reliable tracers of the underlying old stellar population. 

\subsubsection{Standpoint 1: No Bahcall-Wolf cusp in the GC}

If the RG core indeed reflects the intrinsic distribution of the old stars,
then the density profile of the old stellar cusp is genuinely flat.  From this
point of view, one may imply that the Bahcall-Wolf cusp either was disrupted at
some time within the past few Gyr or has never formed in the GC. 

In fact, flat-cored galaxies are ubiquitous in the local universe
\cite{faber97,kormendy13}.  A conventional explanation is that their NSCs are
eroded due to the dynamical evolution of SMBH binaries, pairs of SMBHs that are
repeatedly restored in galaxy centers following galaxy mergers or infalling of
globular star clusters \cite{begelman80,merritt06,kormendy09}. Such a galactic
nucleus has the characteristic that the amount of (initially bound) stars that
have been removed  is correlated with the reduced mass of the SMBH binary
\cite{zier06,zier07,sesana08}.  Similar ideas have been applied to the GC
\cite{baumgardt06,portegies06,matsubayashi07,lockmann08imbh,gualandris12}. A
contrast between a {\it bona fide} Bahcall-Wolf profile and the density profile
in the GC shows that about an amount of $2\times10^5~\msun$ of stars is
currently missing \cite{lang13}. The ``mass deficit'' could have been much
larger in the past, because over a time span of several Gyr, stellar relaxation
may have significantly reduced the core size by gradually recovering the
Bahcall-Wolf profile \cite{merritt10}. To remove this large amount of stars
from the GC, the secondary BH is expected to be massive, probably
$O(10^5)~\msun$ \cite{lang13}. A massive secondary BH could in principle induce
perceivably large perturbations to the kinematics of \Sgr and the surrounding
stars, but so far there is no empirical evidence of these disturbances
\cite{hansen03,yu03,trippe08,gualandris09,chen13}. 
 
An alternative explanation to the non-existence of the Bahcall-Wolf cusp is
that the cusp never has formed. This could be the case if the GC has been built
up in a dissipationless fashion, for example, due to mergers of infalling star
clusters \cite{capuzzo93,capuzzo08}. In this way, stars will be deposited
preferentially at large distances from \Sgr where clusters become
disrupted by the strong tidal field \cite{antonini13a}, and under two-body
relaxation, it will take longer than a Hubble time to develop a Bahcall-Wolf
cusp. Recent numerical simulations demonstrated that consecutive mergers of
about $10$ massive star clusters indeed can build up a NSC with a flat density
profile resembling that in the GC \cite{antonini12,antonini14,perets14}.
However, it is not clear whether the same model could reproduce as well the
large degree of rotation which has been discovered in the central parsec of the
NSC in our Galaxy \cite{becklin68,schodel14,feldmeier14}.

\subsubsection{Standpoint 2: A Bahcall-Wolf cusp in disguise}

On the other hand, one may consider the possibility that the RGs in the GC do
{\em not} trace the underlying old stellar population.  Then, it is possible
that a Bahcall-Wolf cusp, composed of dim main-sequence stars and compact
stellar objects (e.g. neutron stars and stellar-mass BHs), still exists in the
GC. But a peculiar mechanism must be identified that can specifically deplete
the RGs.

One such mechanism which has been proposed in the literature is stellar
collision \cite{genzel96,davies98}. It is more effective in depleting RGs than
other types of stars because RGs have larger sizes, therefore larger
collisional cross sections. Further calculations showed that stellar collision
may explain the deficiency of RGs in the innermost  $1\arcsec$ of the GC
\cite{alexander99}, given the condition that a Bahcall-Wolf cusp composed of
compact stellar remnants does exit.
But further out, because of the lower density of background stars,  stellar
collision becomes less frequent and more difficult to deplete the RGs there
\cite{bailey99,dale09}.

\subsection{Hiding the Cusp during the Fragmenting Past of the Mini
Disk}\label{sec:frag}

Recently, we proposed a new mechanism \cite{pau14}, which does not rely on
hypothetical ingredients, such as a secondary massive BH, a complicated merger
history in the GC, or the pre-existence of a steep Bahcall-Wolf cusp.
Meanwhile, the mechanism can deplete the RGs out to a large distance from \Sgr
that matches observations.  In this mechanism, the culprit for the missing RGs
is the mini disk recently discovered in the GC (see Section~\ref{sec:intro}).
The key idea is that the WR/O stars, the current constituents of the disk, must
have been born a few Myr ago out of massive (e.g. $10^2~\msun$) gaseous clumps
\cite{levin03,nay05,levin07}, and these clumps may destroy the envelopes of the
RGs by repeatedly colliding with them.  As a result, the RGs do not trace the
distribution of the other types of stars.

At the core of the clumpy-disk mechanism is the process called ``ram
stripping''.  A RG ramming into a gaseous medium will gain certain amount of
linear momentum in its outer envelope, and consequently, a fluid element in the
outer envelope will receive a kick velocity. The kick velocity becomes
comparable to the escape velocity at the surface of the RG if the collision
happens, for example, in the central region of a quasar (the most luminous type
of AGN), where the collision velocity increases to $>10^3~\kms$ and the surface
density of the gas medium, mostly likely in a disk configuration called
``accretion disk'', becomes as high as $10^5~{\rm g~cm^{-2}}$.  In such
an extreme condition, each collision with the gas disk will result in a
significant mass loss from the RG, and the entire RG envelope can be removed by
about a dozen collisions  \cite{armitage96}.

The same condition for RG envelope depletion could be met in the GC.  But
first, some modifications to the conventional picture (\cite{armitage96}) have
to be made, because in the GC the accretion disk on average has a very low
surface density, merely $\Sigma_d\sim10^4~\msun/(0.1~{\rm pc})^2\sim200~{\rm
g~cm^{-2}}$, which is inferred from the current stellar constituents of the
mini disk.  Instead, efficient ram stripping is caused by the collisions with
the gas fragments (clumps), which must have existed in the past to give birth
to the observed $O(10^2)$ WR/O stars in the disk.  These clumps are so dense
that the self-gravity is able to overcome the large turbulent pressure,
allowing the clumps to collapse to form stars. Their surface densities are
typically $10^{4-5}~{\rm g~cm^{-2}}$, comparable to the surface densities of
the quasar accretion disks. It has been shown that collisions with RGs
hardly have any effects on such clumps \cite{pau14}.

In the paper outlining the clumpy-disk mechanism \cite{pau14}, we
considered 100 clumps ($10^2~\msun$ each) whose number and spatial distribution
match the observations of the WR/O stars in the mini disk. We have shown that
each RG entering the sphere of radius of $0.1$ pc centered on \Sgr will
experience $2-60$ collisions with different clumps, therefore is prone to
complete removal of its envelope.  The exact number of collisions depends on
the orbital inclination: a RG whose orbital plane is aligned
with the mini disk is more frequently bombarded than a misaligned RG would be.

The efficiency in depleting RGs by the clumpy-disk mechanism is
determined by the following three elements. In particular, (i) and (ii) should
both be effective. Otherwise, RG depletion will be restricted to a small region
very close to \Sgr, where the clumps are the densest, and consequently, the
core will be smaller than what is observed in the GC.

\begin{enumerate}

\item {\bf Non-linear mass loss:}
In the numerical simulations carried out by Armitage et al. \cite{armitage96},
the collision is between a RG of a size of $150~\rsun$ and an accretion disk of
a surface density of $10^{4-5}~{\rm g~cm^2}$. For these parameters, the
collision will efficiently heat the RG, forcing the gas envelope of the RG to
expand to a larger radius. Because of the expansion of the RG, it is suspected
that during the next collision the envelope will lose twice more mass.  Such a
process of mass loss is non-linear. Assuming that the increment of mass loss
during successive collisions is a factor of two, in other words, the non-linear
factor is $f_{\rm loss}=2$, then one will derive that after only $14$
collisions the RG will be completely deprived of its envelope \cite{pau14}.

The degree of the non-linearity ($f_{\rm loss}$) is a crucial parameter of our
model. It is determined by the relative efficiency of the heating (energy gain)
and cooling processes in the RG, which in turn depends very much on the
structure of the RG envelope. Since the most common RG has a radius smaller
than $150~\rsun$ and the energy gain is proportional to the projected area of
the RG,  it is expected that heating will be less effective for these RGs.  On
the other hand, for these typical RGs, cooling will be more important because
of the shorter hydrodynamical timescale. As a result, in a more realistic
situation, the non-linearity factor $f_{\rm loss}$ may be close to unity,
making the RGs more rigid against ram stripping. For example, if $f_{\rm
loss}=1.1$ ($1.01$), the number of collisions that is needed to completely
remove the envelope of a RG increases to $80$ ($530)$. To quantify $f_{\rm
loss}$ for RGs of different sizes, more numerical simulations are needed. 

\item {\bf Longevity of the clumps:}
Unlike the clumps in conventional regions of massive star formation (SF), which are
optically thin and will collapse on a free-fall timescale of $10^3$ yr
\cite{zinnecker07}, the clumps close to \Sgr are extremely opaque due to their
high densities \cite{pau14}. As a result, the heat released from gravitational
collapse cannot escape freely, so the clumps are expected to contract slowly,
on a timescale of several $10^5$ yr.  The elongated lifetimes of the clumps
near \Sgr significantly increase the chance of RG-clump collisions.

The above rather unconventional picture of SF process may be further
complicated if additional physical processes are taken into account, such as
clump collision and gas accretion (\cite{goodman04,nay05,nay06subpc,levin07}).
The clumps shaped by these additional processes are likely to have very
different geometries as well as lifetimes. These  differences may result in a
variation of the probability of collision with the RGs, which deserves further
investigation.

\item{\bf Number of clumps:}
The number of collisions mentioned above should be regarded as a lower limit,
because in the calculation we neglected the contributions from the clumps with
masses $10~\msun$ or smaller, whose number is observationally less well
constrained.  These lighter clumps in general have higher surface densities
(see derivation in \cite{pau14}), therefore will strip more mass from RG during
each collision. Furthermore, the lighter clumps close to \Sgr normally have
longer contraction times, because at birth their sizes are smaller so they
radiatively dissipate the gravitational energy at a much slower rate
\cite{pau14}. The higher surface density and the longer lifetime compensate the
smaller collisional cross section caused by the smaller size, making a light
clump as effective as the more massive ones (e.g. $10^2~\msun$) in terms of
depleting the envelopes of RGs.

So far, it is unclear how many light clumps existed in the mini disk.
Formation of a large amount of light clumps ($\sim10~\msun$ or less) has been
reported in the numerical simulations of the mini disk, especially those
assuming effective gas cooling \cite{nay05,bonnell08,hobbs09,alig11,mapelli12}.
But it is difficult to put the theoretical numbers into observational test,
because the stellar descendants of the light clumps are normally faint, under
the detection limit of the current telescopes \cite{bartko10,do13,lu13}.  There
have been some hints, from the X-ray emission in the GC \cite{NaySun05} and
from the stellar dynamics of the mini disk \cite{nay06weigh,alexander07},
suggesting that the total mass of the ``unseen'' stellar population is small.
Deeper photometric and spectroscopic observations will help to better
constrain the abundance of these clumps. 

\end{enumerate}

There are at least two possibilities to observationally test the clumpy-disk
scenario for RG depletion.  (1) Today, the cores released from the destructed
RGs are still populating the GC, in the central $0.1$ pc region. Detection of
them will strongly support the clumpy-disk scenario.  These cores initially
have high effective temperature (because of the thin envelopes) and are
expected to be as luminous as their progenitor RGs \cite{davies05}.  Therefore,
photometrically they resemble main-sequence stars of $3-4~\msun$, and deep
spectroscopy is needed to reveal their real masses. (2) The surviving RGs,
especially those most close to \Sgr, should preferentially lie on the orbital
planes that are inclined with respect to the mini disk.  Otherwise, they cannot
avoid large number of collisions with the clumps and their envelopes should
have been depleted.

\section{Randomizing S-stars}\label{sec:Sstars}

\subsection{More than One Paradox}\label{sec:paradoxes}

Discovery of the S-cluster, and the observational follow-up in the past
decade, has challenged the dynamical model of the GC with a series of
dilemmas.

\begin{enumerate}

\item {\bf Paradox of youth:}
It is generally believed that SF will be prohibited in the vicinity of \Sgr,
e.g. within a distance of $1~\arcsec\simeq0.04$ pc, because of the violent
environment, such as the strong tidal force, large turbulent velocity, and
strong radiative background \cite{morris93}. Only old stars have had long
enough time (several Gyr) to diffuse into the central region due to various
stellar dynamical processes \cite{pau12}. This theoretical expectation,
however, apparently contradicts the existence of the young (life expectancy of
$6-200$ Myr) S-cluster inside a region as close as $<1\arcsec$ from \Sgr. This
contradiction imposes a paradox which is termed ``the paradox of youth''
\cite{morris93,ghez03}.

This paradox can be (and has been) resolved because the S-stars can be
produced further out and later brought in by some fast dynamical process (see
\cite{genzel10} for a review).  One such process is ``binary separation''. A
star binary formed far away from \Sgr could be scattered to a highly eccentric
orbit, such that at the pericenter the binary will be tidally separated by the
SMBH, leaving one star, which could be a B star, tightly bound to the SMBH and
the other flying away with escape velocity
\cite{hills88,hills91,gould03,gin06}.  This binary-separation model can explain
the number and spatial distribution of the majority of the B stars within
a distance of about $1$ pc from \Sgr \cite{perets07,perets10,madigan14}. The
other possibility is ``disk migration''.  A young star formed in the mini disk
could lose its orbital energy and angular momentum by tidally interacting with
the gas in the disk, then it will migrate toward the disk center
\cite{levin07}. In this model, a successful transportation of S-stars into the
central $1~\arcsec$ region requires the existence of gas disk there, although
currently there is no direct evidence supporting this prerequisite.

\item {\bf Paradox of randomness:}
Further analysis of the kinematics of the S-cluster revealed a second paradox.
This is because in the binary-separation model, the captured stars initially
have very high eccentricities, about $0.93-0.99$ (see the original work of
\cite{hills91} and \cite{pau12} for a review), while in the disk-migration
model, the stars in the disk normally have near-circular orbits \cite{nay07}.
Both eccentricity ranges are too narrow to match the observed ``super-thermal''
distribution $dN/de\propto e^{2.6}$ of the S-stars \cite{gillessen09}, where
$e$ is the orbital eccentricity.

This discrepancy has led to the necessity of including stellar-relaxation
processes in the dynamical models, to randomize the angular momenta of S-stars,
therefore broaden the initially narrow distribution in the eccentricity space.
Two dynamical processes, which are inherent to any star-cluster system, have
been considered first.  One is the incoherent (random) scattering between two
stars (two-body relaxation \cite{bt08}), and the other is the coherent torquing
between the stars on near-Keplerian orbits, known as (scalar) resonant
relaxation, or (scalar) ``RR'' \cite{rauch96,rauch98}.  Numerical simulations
including both processes showed that at least $20-25$ Myr is needed to recover
the observed eccentricity distribution if S-stars are injected {\em
continuously} through binary separations
\cite{perets09,madigan11,antonini13,zhang13,hamers14}.  If S-stars are produced
by disk migration, a timescale longer than $100$ Myr will be needed, since the
near-circular orbits preferentially have larger angular momenta
\cite{antonini13}.  These timescales, however, are longer than the life
expectancy of one of the S-stars, S2/S0-2, which is about $6-10$ Myr
\cite{ghez03,eis03,martins08}.

The situation becomes more intractable when one realizes that the
relaxation timescales were derived under the assumption that a dense
Bahcall-Wolf cusp exists in the GC, an assumption inconsistent with
observations (Section~\ref{sec:RGs}). Without a Bahcall-Wolf cusp, the
relaxation timescale is estimated to be much longer than $100$ Myr (e.g.
\cite{antonini13}), then most S-stars will have difficulties evolving within
their lifespans to their current orbital eccentricities. For this reason, the
randomness of S-stars remains a problem.

\item {\bf Paradox of concurrence:}
Even if a Bahcall-Wolf cusp does exist so that paradox ii can be
mitigated, there is yet another paradox.  Occam's razor would favor a scenario
that the young stars (B, O, and WR stars) in the GC have the same origin, e.g.
in one single SF episode they were all produced. Three observations seem to
support this scenario: (1) among the observed B stars the one with the shortest
life expectancy ($6-10$ Myr for S2) seems to be coeval with the WR/O stars
($6\pm2$ Myrs old), (2) the luminosity function (probability distribution as a
function of stellar luminosity) of the entire young stellar population is
consistent with a single SF episode $2.5-6$ Myrs ago \cite{lu13}, and (3)
recent analysis of the most complete data set no longer shows large discrepancy
in the kinetic properties of the B and the WR/O stars \cite{yelda14}.

Despite its simplicity, the coeval scenario conflicts with the theoretical
prediction based on analysis of the relaxation timescales, that many B stars in
the S-cluster should be older than $20-25$ Myr. The root of this last conflict
clearly is entangled with that of paradox ii. It also has difficulties
explaining the apparent spatial segregation between B and WR/O stars: not a
single WR/O star is inside the central $1\arcsec$ region but dozens of B stars
are \cite{alexander11}.

\end{enumerate}

\subsection{A Rapidly Evolving Region Created by the Disk}\label{sec:rer}

The above paradoxes highlight the incompleteness of the simplest model of the
GC -- it is still missing a mechanism that can efficiently randomize the
angular momenta of the S-stars within $10$ Myr.  One possible solution is to
introduce a massive perturber in the central $1\arcsec$ region of the GC.  For
example, the perturber could be an intermediate massive black hole (IMBH) in
the mass range of $(400\sim4,000)~\msun$ and residing at a distance of about
$0.01\arcsec\sim1\arcsec$ from \Sgr \cite{merritt09,gualandris09}.  However, it
is observationally challenging to confirm the existence of such an IMBH so
close to \Sgr \cite{hansen03,yu03,trippe08,chen13,bartos13}.

Recently, we have put forward a new idea, based on analytical estimates of the
timescales, that the perturber is the progenitor of the currently-observed mini
disk \cite{chen14}.  An axisymmetric structure around a SMBH, such as the mini
disk surrounding \Sgr, will gravitationally torque the orbit of a star, forcing
both its orbital eccentricity and its orbital orientation to change.  The
subsequent secular evolution is known as the ``Kozai-Lidov'' (KL) evolution
(\cite{kozai62,lidov62,naoz13}), named after the two dynamists who in the 1960s
pioneered the analysis of such a process in planetary systems. In the
following, we will show that because of the existence of the mini disk, the
S-stars in the GC are also subject to the KL evolution. 

\subsubsection{The Mini Disk in the Past}\label{sec:NewDisk}

There is a key difference between our model and those earlier ones studying the
disk effect \cite{subr05,lockmann08,lockmann09,chang09,madigan14}. This
difference has led to opposite conclusions on the effectiveness of the KL
mechanism.  The earlier models adopted a disk which is identical to the
currently observed one: it has a total mass of $M_d\simeq10^4~\msun$, a
surface-density distribution of $\Sigma_d(R)\propto R^{-\alpha}$
($\alpha=1.4\sim2$), and most importantly, it extends from an outer radius of
$R_{\rm out}\simeq12\arcsec$ ($\simeq0.5$ pc) to an inner radius of $R_{\rm
in}\simeq1\arcsec$ ($\simeq0.04$ pc) \cite{bartko10}.  In this configuration,
the KL evolution will actually be suppressed, because the ``effective mass'' of
the disk that is causing the KL evolution is exceeded by the mass of the NSC
that is enclosed by the orbit of a test star \cite{chang09}.  Under such
condition, the gravitational potential of the NSC can induce a substantial
Newtonian precession to impair the coherence of the KL interaction.

However, there is no reason to assume that the properties of the mini disk have
not evolved during the past several Myr. Since the SF efficiency is normally
smaller than unity \cite{nay07,wardle08,bonnell08,hobbs09,alig11,mapelli12}, it
is reasonable to consider the possibility that initially the disk was more
massive than $10^4~\msun$, which is the total mass of the WR/O stars currently
inside the disk \cite{paumard06,bartko10}. Furthermore, since the disk
originally was composed of gas, which could shed angular momentum by viscosity,
it was likely to be more extended toward the central SMBH
\cite{levin03,nay05,levin07,wardle14}.

For these reasons, we adopt in our model $M_d=3\times10^4~\msun$ and $R_{\rm
in}=10^{-6}~{\rm pc}$ (innermost circular orbit) as the initial conditions of
the disk, meanwhile we keep $R_{\rm out}=0.5$ pc and $\Sigma_d(R)\propto
R^{-1.4}$ ($\alpha$ from \cite{bartko10}).  In this new configuration, the
central part of the disk within the radius of $R=1\arcsec$ contains a mass of
$\Delta M_d\simeq6,000~\msun$. This ``inner disk'' will play a crucial role in
sustaining the KL evolutions of the S-stars. It is not surprising that today
the inner disk is no longer present.  This is because the gas would have been
consumed by SF or accretion onto the SMBH
\cite{levin03,nay05,levin07,alexander12,wardle14}.  After that, within several
Myr, the stars, if there were any in the inner disk, would have been lifted out
of the disk plane due to a particular consequence of RR, or ``vectorial
RR'', which is to randomize the orientation of the stellar orbits
\cite{hopman06,kocsis11}. 

\subsubsection{Numerical Simulations}\label{sec:NumSim}

We have run numerical simulations to test the idea that the extended disk can
induce KL evolutions of the S-stars. In these simulations, the gravitational
potential is a combination of three components: (1) the {\em extended} disk
(Section~\ref{sec:NewDisk}), (2) the central SMBH with a mass of
$\mbh=4\times10^6~\msun$, and (3) a spherical stellar cusp with {\it
shallow} density profile $\rho\propto r^{-1.3}$ (Equation~[4] from
\cite{genzel10}), representing the observed NSC.  The disk in our simulation is
represented by 150 particles, each with a mass of $200~\msun$ and set on a
circular orbit.  This set up accounts for the clumpiness of the fragmenting
disk.  For the black hole, we adopt the  pseudo-Newtonian potential
\cite{paczynsky80}. This potential allows us to investigate the impact of the
GR precession on the KL evolution (e.g.  \cite{chen08,chen11,wegg11}), which is
expected to be important when $e$ becomes large. In our model, the mass of the
NSC enclosed by a sphere of radius $1\arcsec\simeq0.04$ pc is about
$7,000~\msun$. This value is comparable to the mass of the ``inner disk'',
which is $6,000~\msun$, as mentioned above.

\begin{figure}
\centering
\mbox{\subfigure{\includegraphics[width=0.5\textwidth]{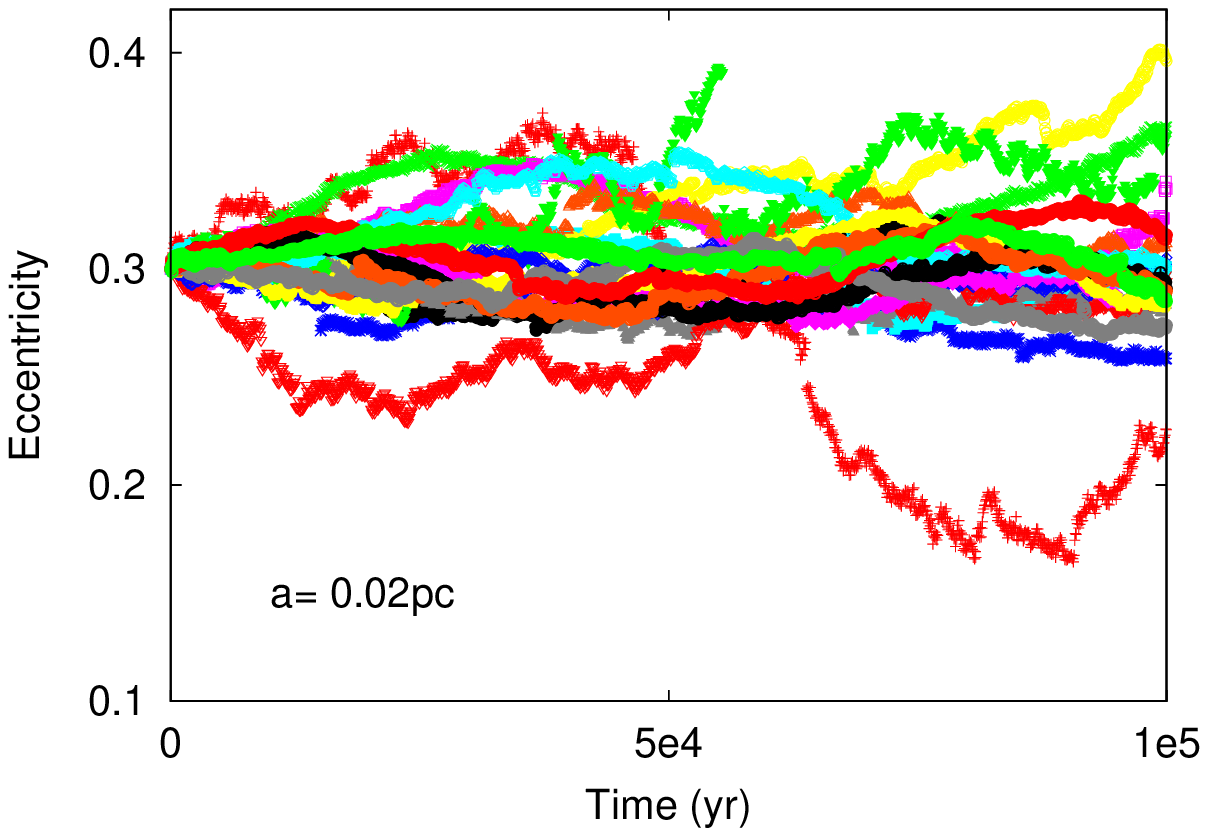}
\quad
\subfigure{\includegraphics[width=0.5\textwidth]{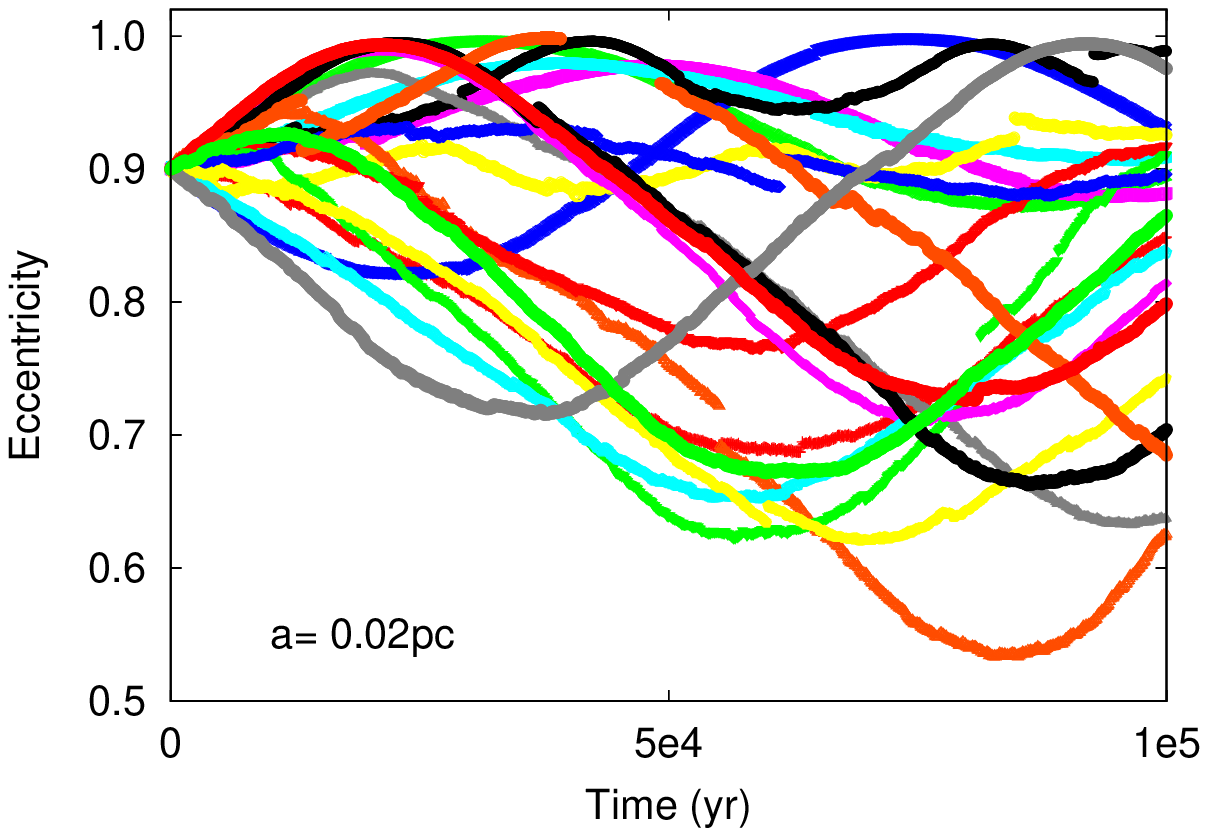} }}}
\caption{Examples of the Kozai-Lidov evolution excited by the {\it extended}
disk. Test stars initially have the same semimajor axis, $a=0.02$ pc.  The
initial orbital eccentricity is either $e_0=0.3$ (left panel), representing the
S-stars formed in the disk-migration model, or $e_0=0.9$ (right panel),
representing those in the binary-separation model.  Different curves have
different orbital inclinations (lines different colors) relative to the plane
of the disk.} \label{fig:kl} 
\end{figure}

Figure~\ref{fig:kl} shows the typical evolutions of the test stars with the
same initial semimajor axis $a=0.02$ pc, but different orbital eccentricities,
i.e. $e=0.3$ and $0.9$. Each curve is derived from a numerical integration of
the equations of motion of a test particle.  Initially, the orbital plane has a
random orientation with respect to the disk plane.  The right panel of
Figure~\ref{fig:kl} shows that the evolution of the orbital eccentricity is
mostly cyclic. This oscillatory behavior has been well studied, and can be
attributed to the two integrals of motion of the system (see \cite{ivanov05}
for examples).  The curves in the left panel of Figure~\ref{fig:kl} show more
irregularities.  This is partly because the Newtonian precession is more
substantial for a test star with lower eccentricity --- more mass from the NSC
is enclosed by the orbit of the star. More importantly, it is due to the longer
oscillation period --- it takes longer time for the disk torque to alter the
larger angular momentum (smaller $e$). To be more quantitative, the KL
formalism (e.g. \cite{ivanov05,chang09,naoz13}) predicts that a test star will
complete a KL cycle on a timescale of
\begin{equation}
t_K(a,e)\sim l(\mbh/\Delta M_d)P(a),\label{eqn:tK}
\end{equation}

\noindent
where $P(a)$ is the orbital period of the test star whose orbital semimajor
axis is $a$, and $l=\sqrt{1-e^2}$ is the angular momentum of the test star
normalized by the angular momentum of a circular orbit with the same $a$.  The
linear dependence on $l=\sqrt{1-e^2}$ stems from the fact that the disk torque
is coherent during one KL cycle. In our model and given $a=0.02$ pc, $t_K$ is
about $3\times10^5\sqrt{1-e^2}$ yr. The results from our numerical simulations
are consistent with this analytical prediction. 

\begin{figure}
\centering
\includegraphics[width=\textwidth]{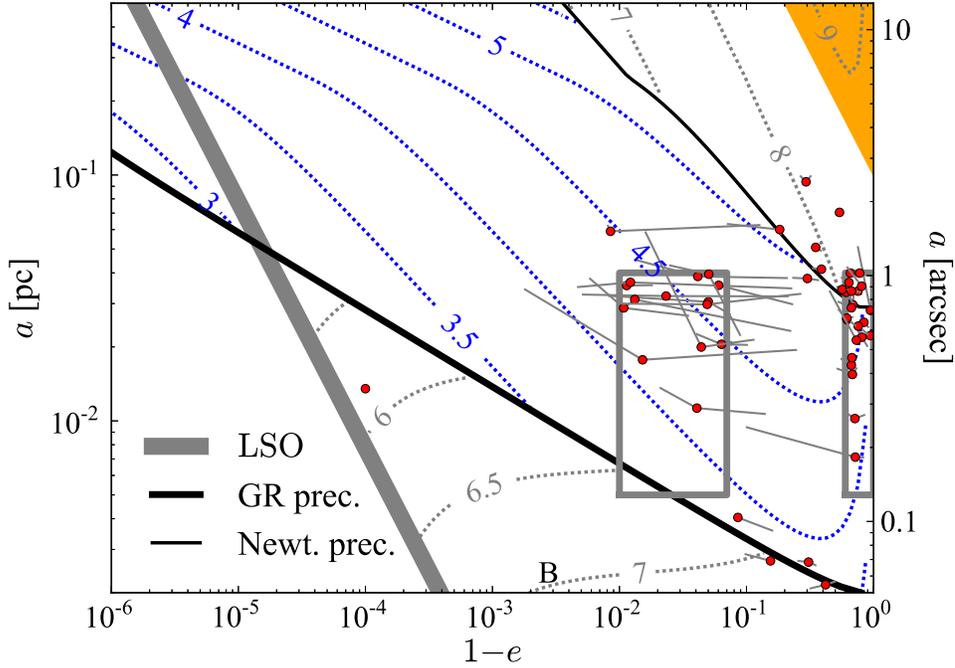}
\caption{Evolutions of the test stars in the phase space of $1-e$ and $a$.  The
region delimited by the two black lines is the RER. In it, the evolution is
dominated by the KL mechanism, and the blue dotted isochrones are associated
with the logarithms of the KL timescales.  Above the thin black line, the KL
evolution will be quenched by Newtonian precession. Below the thick black line,
the quenching is due to GR precession. In these two regions outside the RER,
the evolution is dominated by either two-body relaxation or RR, and the gray
dotted isochrones are associated with the logarithms of the two-body-relaxation
or RR timescales, whichever is shorter.  The thick gray line on the left-hand
side corresponds to the last stable orbit (LSO) around Sgr A∗.  The small
orange triangles at the top right corner depicts the location of the red giants
in the GC \cite{merritt10}. The two gray boxes depict the expected birth places
of S-stars in the binary-separation and migration-in-disk models (also see
\cite{antonini13}). The red dots mark the initial loci of the test stars in our
simulations, and the gray lines show the extensions of the KL evolutions during
$3\times10^5$ yr.  }\label{fig:rer} 
\end{figure}

Figure~\ref{fig:rer} summaries more results from our numerical simulations.  It
is depicting the $(1-e)-a$ phase space of the stars at the innermost
$10\arcsec$ region of the GC.  In this diagram (see \cite{chen14} for
derivation), a star inside the region delimited by the two black lines is
expected to be driven by the KL torque, therefore will evolve, mostly
horizontally, on the KL timescale. The blue contours inside this region
indicate the evolution timescales, defined by $|(1-e)/\dot{e}|$.  We call this
part of the phase space the ``rapidly evolving region'' (RER), because the
evolution timescales are much shorter compared to the timescales outside,
indicated by the gray contours. Outside the RER, the evolution timescales are
determined by either two-body relaxation or RR, whichever is shorter. Here, the
KL evolution will be quenched by either the Newtonian (the upper-right part of
the phase space) or the GR precession (the lower-left part).

It is worth noting that our simulations, by construction, cannot capture the
physical processes such as the two-body relaxation and (scalar) RR. But as can
be seen in Figure~\ref{fig:rer}, in the central $10\arcsec\simeq0.5$ pc of the
Galaxy, both relaxation processes are operating on a timescale of
$\sim(1-10^3)$ Myr (depending on $e$), much longer than the KL timescale.
Therefore, on the timescale relevant to our simulations, which is shorter than
$1$ Myr, the effects of two-body relaxation and (scalar) RR are negligible.
This justifies the omission of the two-body and resonant relaxation processes
in our numerical simulations. 

The gray boxes in Figure~\ref{fig:rer} mark the birth places of the S-stars
predicted by the binary-separation and disk-migration models.  Inside them, the
red dots show the initial loci of the test stars in our simulations. Their
initial semimajor axes are randomly sampled according to the observed density
profile of $\rho\propto r^{-1.3}$ for the NSC \cite{genzel10}. The other red
dots outside the gray boxes represent the test stars which are selected to
demonstrate the typical evolutions close to and outside the boundaries of the
RER.  For each test star, initially we assign a random orbital inclination, and
then we numerically integrate the orbit for $3\times10^5$ yr. On a timescale
longer than $3\times10^5$, the effect of vectorial RR is expected to be
important (see below).  Throughout the integration, we have recorded the
minimum and maximum eccentricities of the orbit. In Figure~\ref{fig:rer}, the
gray lines pointing away from the red dots indicate the loci of the tests stars
when they obtain their extreme eccentricities.  We  note that in our
simulations the $a$ of a test star is not conserved, because the gravitational
potential is varying in time due to the lumpiness of the disk.

Figure~\ref{fig:rer} clearly shows that on a timescale of $3\times10^5$ yr, the
stars inside the RER have moved prominently in the horizontal direction (in
angular-momentum space), while those lying well outside do not move as much.
This result confirms our idea that the KL evolution will be excited by the {\em
extended} mini disk.  Furthermore, the KL evolutions are more prominent for the
stars formed in the binary-separation model (left gray box) than in the
disk-migration one (right gray box). We have mentioned before that this
difference is mainly due to the linear dependence of $t_K$ on $\sqrt{1-e^2}$.

On a timescale longer than $3\times10^5$, we expect the test stars in the RER
to move further away from their initial locations.  This further ``mixing'' in
the angular-momentum space is driven by vectorial RR: it perturbs the
orbital inclination $\theta$ of a test star \cite{rauch96}, and by changing the
integrals of motion of the KL formalism which are functions of $\theta$ and $e$
\cite{ivanov05}, the perturbation will propagate into the amplitude of the KL
oscillation. Such process is not captured by our scattering experiments,
because vector RR is not implemented.  According to the latest calculation
\cite{kocsis14}, the timescale for vectorial RR to change $\theta$ is
proportion to $P(a)$ and $\sqrt{1-e^2}/(1.05-0.3e)$, and in our model with
$\rho\propto r^{-1.3}$, it is shorter than $6$ Myr for a star at $a<1\arcsec$
(also see \cite{chen14}). This indicates that the KL mechanism, combined with
the vectorial RR, could result in a rapid randomization of the S-stars inside
the central $1\arcsec$ region of our Galaxy.
 
\subsubsection{Further Test of the RER Scenario}

At this point, one could wonder what the distribution function ($dN/de$)
would be like if S-stars are randomized by the RER.  Some hints can be found in
the analytical formulae of KL evolution. In the KL formalism, the time $\Delta
t$ spent by a test star in a particular eccentricity interval $e\sim e+\Delta
e$ is proportional to $e\Delta e/\sqrt{1+e^2}$ (e.g. \cite{ivanov05}).  The
term $1/\sqrt{1+e^2}$ is inversely proportional to the angular momentum of the
star, and it stems from the very fact that an orbit with smaller angular
momentum reacts more quickly to the KL torque. The additional factor $e$
ensures that the axisymmetric torque does not change $e$ of a star on circular
orbit, because the two sections of the orbit on the opposite sides of the disk
plane will receive equal amount of torque. Finally, we can derive
$dN/de=dt/de=e/\sqrt{1+e^2}$, which follows from the fact that in a completely
mixed system the number of stars $\Delta N$ in the interval $e\sim e+\Delta e$
is proportional to $\Delta t$.  As has been shown in \cite{chen14}, the above
distribution function is steeper than a thermal one, due to the additional term
$1/\sqrt{1+e^2}$, and it naturally and faithfully recovers the super-thermal
distribution observed in the S-cluster.

The major caveat in the analytical approach is the assumption of a static disk.
In fact, the timescale for a test star to ``mix'' with other stars in the RER
is comparable to the vectorial RR timescale, the timescale on which the
relevant part of the disk is evaporating.  This last condition means that the
disk torque is diminishing as the test star evolves. The efficiency of mixing
in this situation deserves further investigation, and a self-consistent
treatment of the time-dependent background may require numerical N-body
simulations.

In the future if dimmer, older (than B-type) stars can be discovered in the GC,
there would be an alternative, empirical approach of testing the RER scenario.
Since old and young stars in the central $1\arcsec$ region respond
indifferently to the disk torque, in principle they should obtain the same
distribution function by now.  Without the RER, however, the old stellar
population may have had time to relax but not so much for the young stars
(Section~\ref{sec:paradoxes}). Therefore, the orbital eccentricities of the old
stars inside $1\arcsec$ contain crucial information about whether or not the
RER exists. 

\subsubsection{Advantages of the RER Scenario}

One attractiveness of the RER scenario is that it offers a single, unified
resolution to paradoxes ii and iii
(Section~\ref{sec:paradoxes}).

\begin{itemize}

\item Before explaining how the RER scenario resolves the other paradoxes, it
is worth emphasizing that it alone does not resolve paradox i; mechanisms such
as binary separation and disk migration are needed to form the S-cluster in the
first place.

\item The RER shortens the relaxation timescale of the S-cluster to several
Myr.  Most importantly, the hypothetical Bahcall-Wolf cusp is no longer a
prerequisite --a stellar background with shallow density profile, as has been
observed in the GC, would suffice.  This stellar distribution is fully
compatible with the non-detection of a cuspy profile in the distribution of
RGs.

\item Because of RER, the B stars can be coeval with the WR/O stars.  Earlier
models without RER have difficulties conforming with this proposition: if all B
stars formed simultaneously with WR/O stars about 6 Myr ago, two-body
relaxation and RR will fail to recover the observed distribution of $e$.

So far, one puzzle remains. If B stars are coeval with WR/O stars, what is the
difference that causes the difference in the spatial distribution of these two
types of stars?  This question leads us to the following section.

\end{itemize}

\section{Depleting the Innermost WR/O Stars}\label{sec:WROstars}

\subsection{An Inversed Mass Segregation}\label{sec:ims}

In a close encounter between two stars, the principle of energy equipartition
\cite{bt08} predicts that the less massive star will eventually acquire a
velocity that is greater than the velocity of the more massive star.  The
difference in velocity will lead to a spatial segregation of stars: the more
massive population will sink toward the center of the potential well, meanwhile
the lighter population will be expelled further out.  However, the stellar
distribution in the GC does not conform to this straightforward prediction.  In
the GC, the B stars ($<20~\msun$) are populating the innermost $1\arcsec$
region, constituting the S-cluster, but the more massive WR/O stars
($>20~\msun$) lie exclusively further out. This phenomenon is known as the
``inversed mass segregation'' \cite{alexander11}.

Further analysis of the kinematic data of the WR/O stars revealed several
peculiar signatures, which have inspired some ideas of solving the
inverse-mass-segregation problem. But so far none of the approaches is
conclusive.

\begin{enumerate}

\item Longterm monitoring of the stellar orbits revealed that
roughly-speaking half of the GC WR/O stars are residing in one (possibly
two) disk structure(s) \cite{levin03,paumard06,tanner06,lu09,bartko09}. It is
believed that these stars are the descendants of an accretion disk
\cite{levin03,nay05}.  Today, the stellar disk truncates at a distance of about
$0.8\arcsec$ from \Sgr, and if extrapolated to a distance smaller than
$0.8\arcsec$, there would be about $30$ more WR/O stars \cite{bartko10}.  These
30 WR/O stars are not expected to remain in the disk plane, because during
their ages of several Myr they will be lifted out of the disk plane by
vectorial RR (see Section~\ref{sec:Sstars} for description). Therefore, the
truncation of the stellar disk at a radius of $0.8~\arcsec$ is not surprising.
But the puzzle is that RR does not change the semimajor axes of stellar orbits,
so the 30 WR/O stars, though no longer associated with the disk plane, should
remain inside the central $1\arcsec$ region of our Galaxy.

To explain the fact that not a single WR/O star is observed in the central
$1\arcsec$,  it has been suggested that fragmentation, which is a necessary
condition for SF, may be suppressed at the central part of the previously
gaseous disk \cite{nay05,nay06subpc,levin07,wardle14}. However, even though SF
was indeed prohibited so that initially there was a hole in the mini disk, the
hole would have been refilled of stars within a short timescale of several Myr
due to the tidal interactions between the WR/O stars and the gas
\cite{levin07,griv10}, or due to the mutual interactions between the WR/O stars
themselves \cite{madigan09,gualandris12disk,subr14}. Because of these refilling
mechanisms, it is difficult to understand the sharp inner edge in the spatial
distribution of the WR/O stars.

\item The remaining half of the WR/O stars do not appear to associate
with the disk structure(s) today, and their spatial distribution also exhibits
an cutoff at about $1\arcsec$ from Sgr A∗ \cite{bartko10,San14,yelda14}.  Such a
distribution has been interpreted in the context of the binary-separation
scenario \cite{perets07,perets10,alexander11}. In this scenario, in-situ SF is
prohibited in the central $1\arcsec$ region around \Sgr, and the stars there
were produced solely by the binary-separation mechanism
(Section~\ref{sec:paradoxes}).  Numerical calculations showed that the binaries
supplied by this mechanism are mostly from a distance of about $5$ pc from
\Sgr. This radial range coincides with the region where the abundance of WR/O
binaries appears to be negligible.  

The matter recently becomes more uncertain. On one hand, as mentioned above,
theoretical studies on the dynamics of the mini disk predict a substantial
feeding rate of the disk stars to the vicinity of \Sgr
\cite{lockmann08,lockmann09,madigan09,gualandris12disk,subr14}. On the other,
more WR/O binaries have been discovered in the mini disk
\cite{ott99,martins06,rafelski07,pfuhl14}. These new discoveries point to a
possibly non-negligible supply rate of WR/O stars into the central $1\arcsec$
region around \Sgr, which revives the problem of inversed mass segregation.

\end{enumerate}

\subsection{Depleting the WR/O Stars in the RER}\label{sec:TDE}

Now we know that it is difficult to completely prevent WR/O stars from forming
in the central $1\arcsec$ of the Galaxy. Therefore, it is logical to seek a
mechanism that can efficiently deplete them. In \cite{chen14}, we have pointed
out that the RER can fulfill the task of preferentially depleting the WR/O
stars in the designated region, by driving them to highly eccentric orbits so
that at the orbital pericenters they will be tidally disrupted by the central
SMBH. In this section, we describe in more detail how the RER mechanism would
work. 

\subsubsection{The RER Mechanism}

A star wandering too close to \Sgr will be tidally disrupted by the SMBH. The
critical distance is approximately 

\begin{equation}
r_t\simeq r_*\left(\frac{\mbh}{m_*}\right)^{1/3}
\simeq4\times10^{-6}~{\rm pc}\left(\frac{r_*}{\rsun}\right)
\left(\frac{m_*}{\msun}\right)^{-1/3},
\end{equation}

\noindent
where $m_*$ and $r_*$ denote the mass and the radius of the star.  This
characteristic radius, $r_t$, known as the ``tidal radius''
\cite{hills75,rees88}, sets a critical eccentricity $e_t=1-r_t/a$ for the orbit
of a star, such that the star must have $e<e_t$ to avoid tidal disruption.  The
linear dependence of $r_t$ on $r_*$ hints that stars of larger sizes are more
susceptible to tidal disruption. At first glance, $r_t$ is much smaller
than the orbital semimajor axes of the S-stars, which may give the impression
that tidal stellar disruption is rare. This is not true, as we will show below,
when the RER is taken into account.

\begin{figure}
\centering
\includegraphics[width=\textwidth]{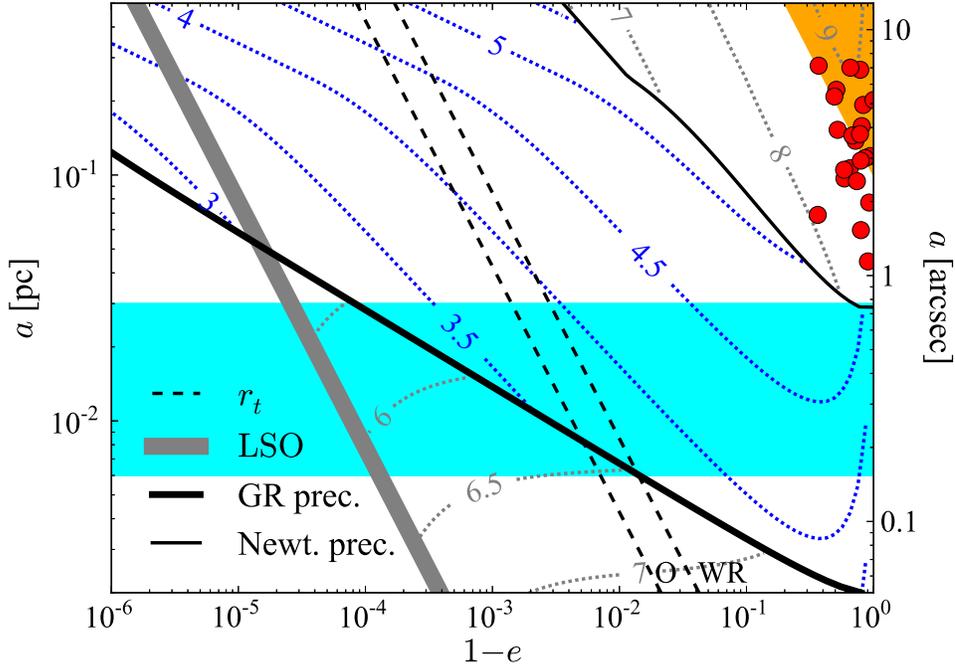}
\caption{Radial range in the phase space where WR/O stars will be tidally
disrupted by \Sgr (cyan-shaded area). The dashed lines labeled with ``O'' and
``WR'' correspond to the tidal radii for, respectively, O and WR stars.  The
red dots represent the subsample of WR/O stars from \cite{paumard06} which are
spectroscopically identified and have 3-D position/velocity measurements.  The
rest lines have the same meaning as those in
Figure~\ref{fig:rer}.}\label{fig:rtWRO} \end{figure}

The function $e_t(a)$ imposes a new type of boundary in the phase space of
$(1-e)$ and $a$, beyond which stars cannot exist.  The two boundaries
corresponding to the tidal radii of WR and O stars are illustrated in
Figure~\ref{fig:rtWRO} as the dashed lines. We adopt
$(m_*,r_*)=(60\msun,80\rsun)$ for the WR stars and
$(m_*,r_*)=(25\msun,60\rsun)$ for the O stars. The other model parameters are
the same as in Figure~\ref{fig:rer}.  The stellar radii adopted here are
somewhat $3-4$ times larger than the values for the main-sequence stars of the
same masses. We choose the larger values because the WR/O stars in the GC are
known to have evolved off the main sequence and entered a transitional
super-giant phase \cite{paumard06,bartko10}.

Figure~\ref{fig:rtWRO} shows that several Myr ago when the mini disk was more
massive and extended, the RER intersected with the tidal-disruption boundaries
at about $a>0.15\arcsec\simeq0.006~{\rm pc}$.  Because of this, no WR/O star
can survive in the cyan-shaded region where $0.15\arcsec<a<0.8\arcsec$: a WR/O
star born in it, whatever $e$ it initially may have, will be transported
horizontally by the RER to the location of $e_t$, and get tidally disrupted.
The timescale for the transportation, as discussed in Section~\ref{sec:NumSim}
as well as in \cite{chen14}, is determined by vectorial RR, and is shorter than
$6$ Myrs for a star at $a<1\arcsec$.  This range of timescale is comparable or
shorter than the ages of the WR/O stars in the GC, therefore it is consistent
with the scenario that the innermost WR/O stars were depleted due to the
existence of the RER.

Figure~\ref{fig:rtWRO} also indicates that WR/O stars may survive at
$a>0.8\arcsec$, to the right of the upper boundary of the RER, where the
evolution timescales are typically longer than $10$ Myr. In fact, those WR/O
stars which are feasible for both spectroscopic identification and 3-D
position/velocity measurement are all detected there (red dots in
Figure~\ref{fig:rtWRO}). In principle, a WR/O star may also survive at
$a<0.15\arcsec$, in the region delimited by the lower boundary of the RER
(thick solid line) and the tidal-disruption radius (dashed line), because there
the evolution timescales are also long.  However, we have not detected WR/O
stars there, maybe because no WR/O was born in that small region: an
extrapolation of the current spatial distribution of WR/O stars would result in
$<1$ star residing at $a<0.15\arcsec$ \cite{bartko10}.

\begin{figure}
\centering
\mbox{\subfigure{\includegraphics[width=0.5\textwidth]{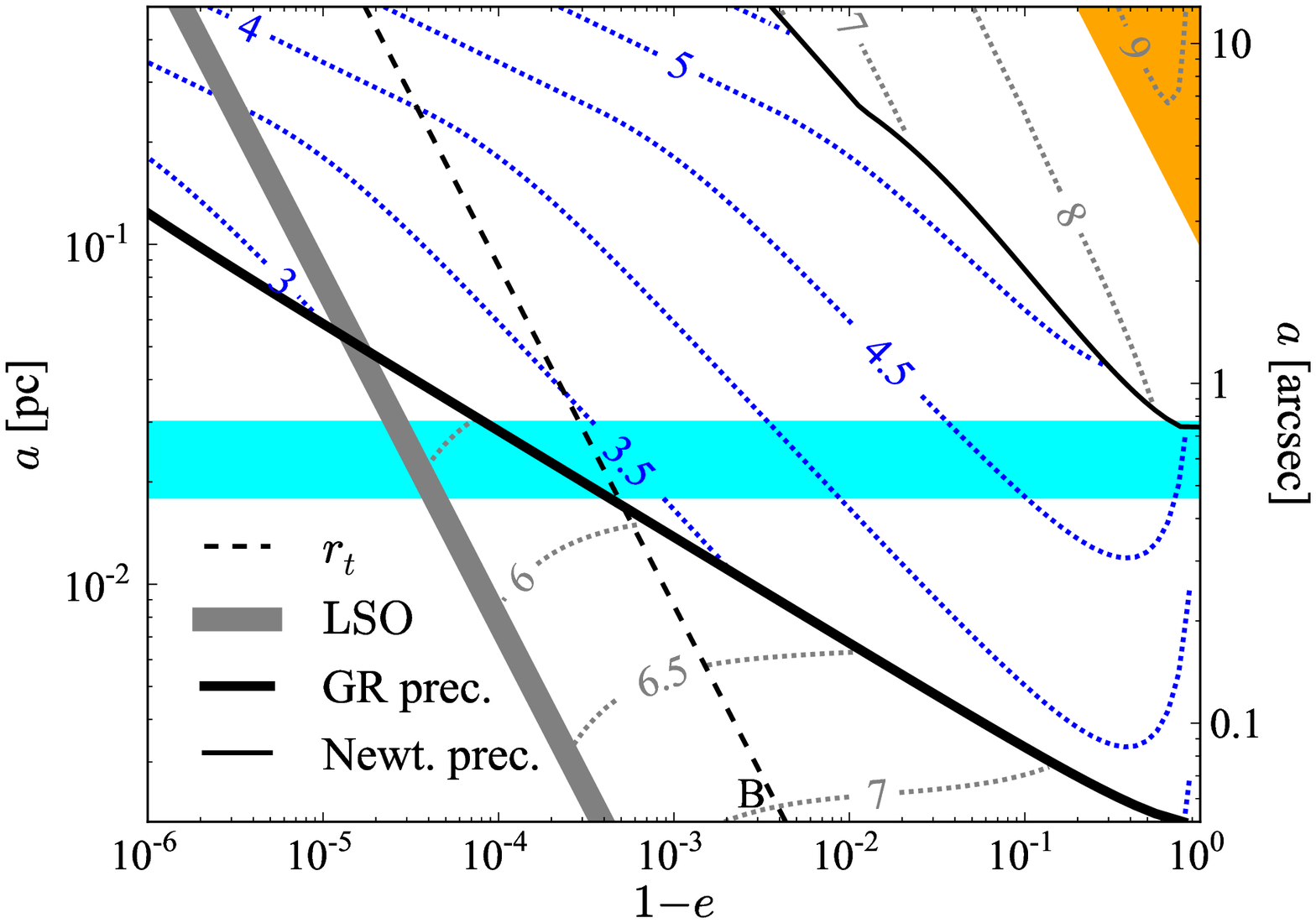}
\quad
\subfigure{\includegraphics[width=0.5\textwidth]{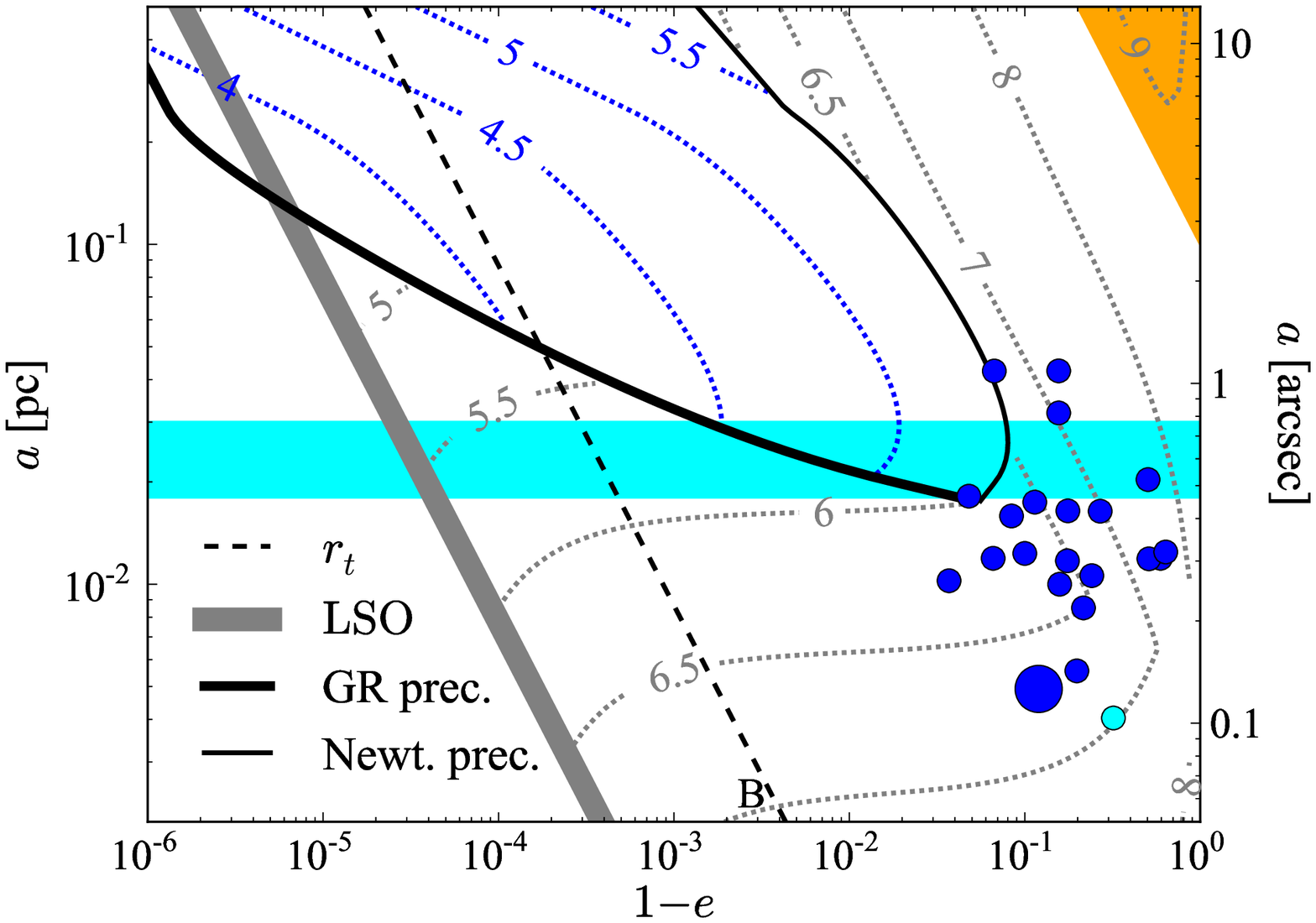} }}}
\caption{{\it Left:} The same as Figure~\ref{fig:rtWRO}, but for tidal
disruption of B stars (assuming $m_*=7~\msun$ and $r_*=4~\rsun$).  {\it Right:}
The same as the left panel, but calculated with the parameters of the present
structures in the GC today, with a disk of mass $10^4~\msun$ and inner radius
at $1\arcsec\simeq0.04$ pc. The dots represent the S-stars not associated with
the mini disk (small-blue \cite{gillessen09}), S2/S0-2 (big-blue, the brightest
S-star \cite{ghez03,eis03}, and S102/S0-102 (small-cyan \cite{meyer12}), the
S-star with the shortest period known.} \label{fig:rtB} 
\end{figure}

Now we proceed to explain why most B stars can be retained in the central
$1\arcsec$ region surrounding \Sgr. It can be understood by looking at
Figure~\ref{fig:rtB}, where the cyan-shaded area depicts the radial range in
the phase space where a B star will eventually be tidally disrupted. This
tidal-disruption region for B stars is much narrower in the radial range than
that of the WR/O stars. This is because a typical main-sequence B-star of mass
$m_*=7~\msun$ has a much smaller radius, $r_*\simeq4~\rsun$, than the radius of
an evolved WR/O star, therefore it has a much smaller $r_t$ too.  This
predicted gap for B stars matches remarkably well the current distribution of
the S-stars (blue dots in the right panel of Figure~\ref{fig:rtB}), supporting
the idea that the RER does play a role in depleting the stars in the GC.  For
stars less massive than $7\msun$, tidal disruption is expected to be less
important, because these stars have an even smaller $r_t$, so the radial range
relevant to tidal disruption becomes diminishing. It will be interesting to
observationally test this last prediction.

\subsubsection{Constraining the Initial  Mass of the Disk}

The condition of retaining the S-stars in the central $1\arcsec$ region around
\Sgr in fact puts an upper limit on the mass of the mini disk. For example, if
we had adopted in our model a larger mass for the initial disk, i.e.
$M_d>3\times10^4~\msun$, the RER in the $(1-e)-a$ diagram would become wider
both vertically and horizontally. This change would result in a broader gap in
the radial distribution of the S-stars, which would readily contradict the
observations.  For this reason, in \cite{chen14} we have constrained the
initial disk mass to be less than $3\times10^4~\msun$.

One uncertainty, which prevents us from pinning down the exact mass of the
disk, is the initial surface density profile of the disk.  We now study the
consequence of this uncertainty.  In the previous calculations, we assumed that
the disk has a power-law density profile, $\Sigma_d\propto R^{-\alpha}$, where
$\alpha=1.4$, which is derived from the present distribution of the WR/O stars
\cite{bartko10}. So if the total mass of the disk is $M_d(R<0.5~{\rm
pc})=3\times10^{4}~\msun$, there will be an amount of $6,000~\msun$ of mass
inside the radius of $1\arcsec$ of the disk (insensitive to $R_{\rm in}$
as long as $R_{\rm in}\ll1\arcsec$). This amount of mass is comparable to the
total mass of the stars residing at the innermost $1\arcsec$ of the NSC,
therefore Newtonian precession does not impair the KL evolution \cite{chang09}
and the RER emerges.

For an arbitrary $\alpha$, the condition of sustaining a RER in the central
$1\arcsec$ of the GC meanwhile retaining most of the B stars in the same region
is, roughly speaking, that the mass in the central $1\arcsec$ of the disk
[$M_d(R<0.04~{\rm pc})$] should be comparable to and not significantly exceed
that in the NSC [$M_c(R<0.04~{\rm pc})\simeq7,000~\msun$]. This condition,
$M_d(R<0.04~{\rm pc})\simeq M_c(R<0.04~{\rm pc})$, leads to an estimate on the
total mass of the mini disk,

\begin{equation}
M_d(R<0.5~{\rm pc})\simeq (50/4)^{2-\alpha}M_c(R<0.04~{\rm pc}) \,\,\,\,\,\,\,\,
{\rm if}~\alpha<2,~{\rm and}
\end{equation}

\begin{equation}
M_d(R<0.5~{\rm pc})\simeq [\ln(50)/\ln(4)]M_c(R<0.04~{\rm pc}) \,\,\,\,\,\,\,\,
{\rm if}~\alpha=2,
\end{equation}

\noindent
which depends only on two observables, $\alpha$ and $M_c(R<0.04~{\rm pc})$.
Here, we do not consider $\alpha>2$, because the total disk mass is divergent
in such an initial condition.  While $M_c(R<0.04~{\rm pc})$ is estimated to be
about $7,000~\msun$ \cite{genzel10}, the value of $\alpha$ is less certain (see
\cite{subr14} and references therein). In our fiducial case with $\alpha=1.4$,
we find $M_d(R<0.5~{\rm pc})\simeq 3\times10^4~\msun$.  Otherwise, if
$\alpha=1$ or $2$ \cite{subr14}, we find $M_d(R<0.5~{\rm pc})$ is approximately
$9\times10^4~\msun$ or $2\times10^4~\msun$. These numbers indicate that
initially the mass of the mini disk is a few percent, or less, of the mass of
the central SMBH.

\section{Summary and Outlook}\label{sec:outlook}

Decades of observations of the GC have revealed three puzzling phenomena, i.e.
the problem of missing RGs, the paradox of youth, and the inversed mass
segregation. They apparently defy the fundamental astrophysical principles.
Explanations of these phenomena have been sought separately, and they often
resort to hypothetical structures that so far have not been detected in the GC.
As a result, the dynamical model of the GC progressively becomes more
complicated.

In this article, we have identified a single, unified resolution to all the
three conundrums, by adding only one ingredient which has been missing in the
previous models --the dynamical effects of the recently discovered  mini disk.
We have shown that the clumps formed during the fragmenting past of the disk
could efficiently strip the envelopes off the RGs in the central $0.1$ pc of
the NSC, therefore duplicate the observed flat spatial distribution of the RGs.
Considering that the disk probably was more massive and also had extended to a
smaller radius in the past, we show that it would impose a RER in the central
$1\arcsec$ ($0.04$ pc) region around \Sgr, in which the S-stars would be driven
to cyclic evolutions (KL evolution) in the eccentricities space.  These induced
KL evolutions could quickly randomize the angular momenta of the S-stars,
therefore can resolve the paradox of youth.  The discovery of the RER also
solves the problem of the inversed mass segregation, because the WR and O-stars
in the central $1\arcsec$ region will intermittently attain ample
eccentricities that will lead to their tidal disruptions by \Sgr. 

Our results point to the picture that several Myr ago our GC was an AGN.  This
picture has been repeatedly invoked to explain an increasing number of
observations
\cite{cheng06,cheng07,ponti10,capelli12,su10,su12,yusef12,BlandHawthorn13}.
The results presented in this article provide an additional evidence, from a
different, dynamics point of view. Further more, our study indicates that the
AGN was powered by gas accretion as well as by tidal disruptions of WR/O stars. 

We also find an upper limit on the total mass of the previously gaseous disk,
which should be several percent of the mass of the central SMBH.  Existence of
such a small disk is consistent with the contemporary scenario that ``small''
SMBHs with $\mbh<10^7~\msun$ are mostly powered by stochastic gas inflow (e.g.
\cite{hopkins06,volonteri13}). In the absence of a preferential direction
for gas supply, the above incoherent mode of mass accumulation is likely to
spin down the central SMBH \cite{king05,lodato06,volonteri07,dotti13}, a
prediction that can be tested by direct measurements of the spin of \Sgr (e.g.
using the event horizon
telescope\footnote{http://www.eventhorizontelescope.org}). However, it is
likely that future observations and theoretical investigations of the GC will
continue to surprise and enlighten us.

\section*{Acknowledgment}

This work has been supported by the Transregio 7 ``Gravitational Wave
Astronomy'' financed by the Deutsche Forschungsgemeinschaft DFG (German
Research Foundation). 

\section*{References}
\bibliographystyle{unsrt}

\end{document}